**Volumetric behaviour of (carbon dioxide + hydrocarbon) mixtures at high pressures**

Johnny Zambrano[a,b], Franklin V. Gómez-Soto[a,b], Daniel Lozano-Martín[a], M. Carmen Martín[a], José J. Segovia[a,*]

[a] TERMOCAL Research Group, University of Valladolid, Engineering School, Paseo del Cauce 59, Valladolid 47011, Spain

[b] Dep. de Petróleos, Escuela Politécnica Nacional, Fac. de Ing. en Geología y Petróleos, Quito, Ecuador

## Abstract

The interest of oil industry in increasing heavy oil production has promoted the use of enhanced oil recovery techniques such as $CO_2$ injection, which produce a decrease of oil viscosity and displacement of heavy oil from reservoir to surface. The design of these processes requires accurate data of densities, viscosities or surface tensions of ($CO_2$ + hydrocarbon) mixtures in order to simulate the behaviour of these mixtures in the reservoir.

An automated Anton Paar DMA HPM vibrating-tube densimeter was used to measure densities of this kind of mixtures, and a new mixture injection system, by means of two syringe pumps, was developed for the densimeter. The equipment operates at high pressure, which is controlled through a back pressure valve and a variable volume cylinder with a stepper motor. The estimated standard uncertainty of the density is ±0.9 kg $m^{-3}$ at temperatures below 373.15 K and pressure range (0.1–140) MPa.

In this paper, the densities of the mixtures ($CO_2$ + *n*-decane), ($CO_2$ + *n*-dodecane) and ($CO_2$ + squalane) are reported at $T$ = (283.15–393.15) K and $p$ = (10–100) MPa.

* Corresponding author. E-mail address: jose.segovia@eii.uva.es (J.J. Segovia).

## 1. Introduction

The growing interest of the oil industry in increasing heavy oil production has caused the use of enhanced recovery processes such as water-flooding, caustic flooding, hydrocarbon injection, micelar-polymer flooding or several thermal methods [1]. Also, carbon dioxide flooding is use for this purpose, which consists in the injection of carbon dioxide in order to reduce oil viscosity and move heavier hydrocarbons from reservoir to surface [2–6]. The design of this enhanced oil recovery process requires low uncertainty data of densities, viscosities or interfacial tensions for hydrocarbon + $CO_2$ mixtures thereby obtaining reproducible simulations of displace- ment processes in reservoirs [7–9].

There are other applications where mixtures $CO_2$ + hydrocarbons are involved, for example, in air conditioning systems where a hydrocarbon is used as lubricant. Some $CO_2$ characteristics are fireproof, inexpensive and non-toxic, which means that it is a natural refrigerant that can replace HFCs with effects on global warming. $CO_2$ requires small amounts of oil to be used in air conditioning systems and it is therefore necessary to know the behaviour of the density of the mixtures [10].

The use of $CO_2$ in supercritical fluid extraction processes has many applications, such as removing the sulfur content of gasoline, and for the design of these processes, accurate knowledge of pVT properties are required [11,12].

## 2. Experimental

### 2.1. Materials

The specifications of the materials used in this research are summarized in Table 1. All of them, which were checked by gas chromatography (GC), were used without further purification.

**Table 1.** Material description.

| Chemical name | Source | CAS registry number | Mass fraction purity (GC) |
|---|---|---|---|
| Carbon dioxide | Air-Liquid | 124-38-9 | ≥0.99998 |
| n-Decane | Sigma–Aldrich | 124-18-5 | ≥0.984 n-Dodecane |
| Carbon dioxide | Air-Liquid | 124-38-9 | ≥0.99998 |
| n-Decane | Sigma–Aldrich | 124-18-5 | ≥0.984 n-Dodecane |

### 2.2. Equipment and measurement procedure description

Density measurements were performed using an Anton Paar DMA HPM vibrating-tube densimeter, which is suitable for the determination of density in the range (0–3000) kg $m^{-3}$ with a resolution of $10^{-2}$ kg $m^{-3}$ [13]. The equipment was automated using a control programme developed in Agilent Visual Engineering Environment (VEE) language [14]. It was calibrated according to the procedure described in the literature [15,16]. When the technique is used for liquid mixtures, they are prepared by weighing and charged manually to the system; now for biphasic mixtures, a different sample injection system was designed.

The new scheme of the experimental equipment can be seen in Fig. 1. There are two injection pumps ISCO Model 260D with a capacity of 265 $cm^3$, one pump is used for charging the hydrocarbon and the other pump contains $CO_2$. Both pumps are jacked and an external thermostatic bath maintains constant the fluid temper- atures: $T$ = 313.15 K for the hydrocarbon and $T$ = 283.15 K for $CO_2$. These pumps are connected by stainless steel pipes and valves to the Anton Paar DMA HPM vibrating-tube densimeter. The pressure control is performed using a step-by-step motor connected to a variable volume and a pressure valve MITY MITE S-91XW. The temperature is controlled by an external bath Julabo F25-HE, which can work in a temperature range from $T$ = (248.15–473.15) K. A thermometer ASL F100 provided with two Pt100 sensors, which were calibrated in our laboratory and traceable to national standards, is use for temperature measurement with an expanded uncertainty ($k$ = 2) of ±20 mK.

A digital manometer, DRUCK DPI 104, allows the measurement of pressure for a pressure range $p$ = (0–140) MPa, with relative expanded uncertainty of ±0.02% ($k$ = 2). The manometer was also calibrated in our laboratory and traceable to national standards.

The mixtures were prepared from known flow rates of the components, which were determined through the injection pumps, and the density of the pure compounds at temperature and pressure of their injection. The molar flow of the components was evaluated using:

$$n_i = \frac{Q_i \rho_i}{M_i} \quad (1)$$

where for each component $i$ in the mixture, $n_i$ is the molar flow, $Q_i$ is the volumetric flow in the injection pump, $p_i$ is the density at the injection temperature and $M_i$ is the molar mass.

The densimeter was filled with a total flow rate of $5 \times 10^{-6}$ m$^3$ min$^{-1}$ using an isobaric process and avoiding $CO_2$ bubbles formation. One pump is filled with $CO_2$ coming from the bottle and kept in liquid phase at 283.15 K and 6.0 MPa (above its saturation pressure of 4.5 MPa). The second pump is filled with the hydrocarbon at the same pressure of 6.0 MPa and 313.15 K to reduce its viscosity. For the filling process, pump flows are programmed to desired compositions, while maintaining a total flow rate of $5 \times 10^{-6}$ m$^3$ min$^{-1}$. Valves which inject fluids in the system are open, first $CO_2$ and immediately hydrocarbon. Both fluids are mixed in a “T” connecting the two pumps. The mixture fills the circuit and, when pressure reaches the value of 6.0 MPa, the filling valve of pressure generator is open and the back-pressure valve allows constant pressure conditions. Flow is circulating until a total volume of 40 cm$^3$ is injected in the system, thus composition and homogeneity of fluid remaining in the vibrating-tube densimeter is guaranteed.

For each sample, ten pressures and eight isotherms are performed, obtaining 80 densities. The whole system is automated, using as criteria for recording, the period stability of the densimeter. In order to check the homogeneity and stability of the mixture, density values were repeated twice at one pressure for each isotherm, the agreement of them was better than the measurement uncertainty. The expanded uncertainty ($k$ = 2) for the density measurements was ±1.8 kg m$^{-3}$ [13]. The composition uncertainty depends on the molar fraction of the components and these values are reported in Table 2.

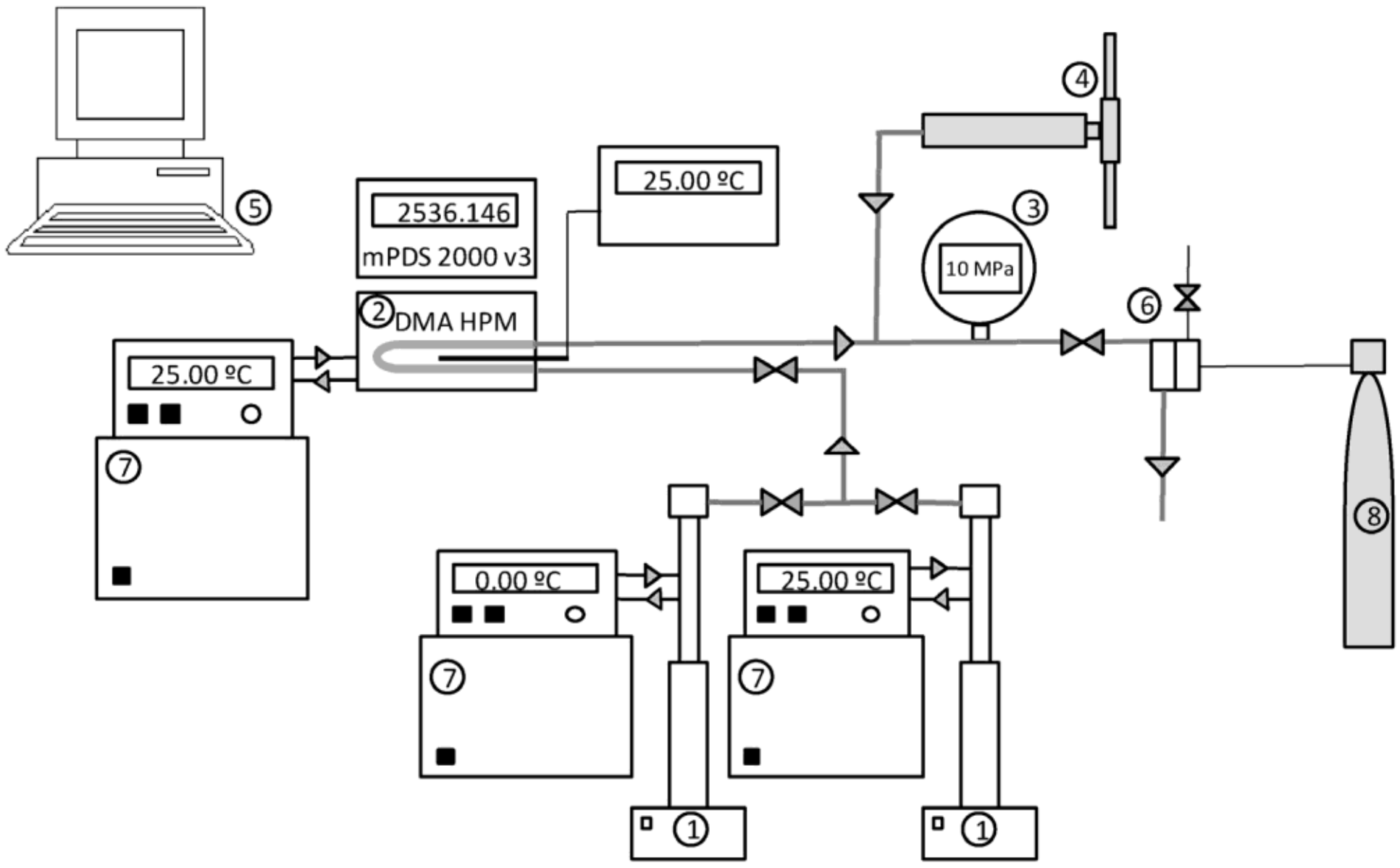

**Fig. 1.** Scheme of the experimental assembly: (1) injection pumps, (2) vibrating-tube densimeter, (3) pressure gauge, (4) automatic pressure control system, (5) computer, (6) Myte Mite valve (pressure controller), (7) thermostatic baths, (8) compressed gas cylinder.

**Table 2.** Expanded uncertainty $U(x)$ ($k$ = 2) of the composition for mixtures {CO2 (x) + hydrocarbon (1 − x)}.

| $x$ | 0.1000 | 0.2000 | 0.3000 | 0.4000 | 0.5000 | 0.6000 | 0.7000 | 0.8000 | 0.9000 |
|---|---|---|---|---|---|---|---|---|---|
| $U(x)$ | 0.0017 | 0.0021 | 0.0026 | 0.0029 | 0.0030 | 0.0028 | 0.0025 | 0.0019 | 0.0010 |

## 3. 3. Experimental results and discussion

The densities of ($CO_2$ + *n*-decane), ($CO_2$ + *n*-dodecane) and ($CO_2$ + squalane) mixtures are reported at eight isotherms in the range $T$ = (283.15–393.15) K and ten pressures in the range $p$ = (10–100) MPa.

The {$CO_2$ (1) + *n*-decane (2)} mixture is proposed as a test system for the validation and development of the two-phase experimental technique, four compositions were measured and the experimental densities are listed in Table 3.

These experimental results were compared with those found in the literature [10,11,17,18] obtaining good agreement. In Fig. 2 can be seen that the relative deviations from the literature are distributed around zero without any trend, although they increase with pressure; the mean absolute deviations obtained in comparison with literature were: 0.5% [10]; 0.4% [11]; 0.6% [17] and 0.7% [18].

The mixture has immiscibility, characterized by a lack of stability in the densitometer reading period at higher temperatures and low pressures, as shown in Fig. 3. The immiscibility conditions are remarked in Table 3.

**Table 3.** Experimental density, $\rho$ (kg m$^{-3}$), for the mixture {($x_1$) $CO_2$ + (1 − $x_1$) *n*-decane} at different pressures and temperatures[a]. IC stands for inmiscibility conditions.

| *T*/K | 283.15 | 293.15 | 313.15 | 323.15 | 333.15 | 353.15 | 373.15 | 393.15 |
|---|---|---|---|---|---|---|---|---|
| *p*/MPa | $\rho$/(kg m$^{-3}$) | | | | | | | |
| $x_1$ = 0.3011 ± 0.0026 | | | | | | | | |
| 10.0 | 766.1 | 757.6 | 739.9 | 731.1 | 722.3 | 703.9 | 684.7 | 665.1 |
| 20.0 | 774.4 | 766.5 | 750.1 | 741.9 | 733.7 | 717.0 | 699.7 | 682.8 |
| 30.0 | 782.0 | 774.5 | 759.1 | 751.3 | 743.7 | 728.1 | 712.4 | 696.9 |
| 40.0 | 789.0 | 781.9 | 767.2 | 759.9 | 752.7 | 738.0 | 723.5 | 709.0 |
| 50.0 | 795.6 | 788.6 | 774.7 | 767.5 | 760.6 | 746.9 | 733.0 | 719.3 |
| 60.0 | 801.5 | 794.9 | 781.5 | 774.7 | 768.2 | 754.8 | 741.7 | 728.8 |
| 70.0 | 807.3 | 800.8 | 787.8 | 781.4 | 775.0 | 762.3 | 749.8 | 737.4 |
| 80.0 | 812.6 | 806.4 | 794.0 | 787.7 | 781.6 | 769.6 | 757.4 | 745.5 |
| 90.0 | 818.0 | 811.7 | 799.6 | 793.6 | 787.8 | 776.1 | 764.1 | 753.0 |
| 100.0 | 822.8 | 816.9 | 805.0 | 799.2 | 793.3 | 782.0 | 770.6 | 759.7 |
| $x_1$ = 0.6052 ± 0.0028 | | | | | | | | |
| 10.0 | 794.8 | 783.2 | 759.3 | 747.3 | 734.5 | 707.0 | 672.6 | IC |
| 20.0 | 806.6 | 796.3 | 774.8 | 764.5 | 753.5 | 729.8 | 704.8 | IC |
| 30.0 | 817.2 | 807.8 | 788.4 | 779.3 | 769.5 | 748.0 | 725.3 | 703.3 |
| 40.0 | 826.9 | 818.2 | 800.8 | 792.5 | 783.4 | 763.9 | 743.5 | 723.1 |
| 50.0 | 835.7 | 827.6 | 811.8 | 804.2 | 795.8 | 777.5 | 759.2 | 740.2 |
| 60.0 | 843.9 | 836.6 | 822.0 | 814.8 | 807.1 | 789.7 | 773.1 | 755.5 |
| 70.0 | 851.7 | 844.8 | 831.4 | 824.7 | 817.2 | 800.8 | 785.5 | 769.2 |
| 80.0 | 859.1 | 852.6 | 840.1 | 833.9 | 826.7 | 811.3 | 796.9 | 781.7 |
| 90.0 | 866.3 | 860.1 | 848.2 | 842.2 | 835.6 | 820.5 | 806.9 | 792.9 |
| 100.0 | 872.8 | 867.1 | 856.0 | 850.0 | 843.6 | 829.4 | 816.4 | 803.2 |
| $x_1$ = 0.7970 ± 0.0019 | | | | | | | | |
| 10.0 | 837.3 | 818.4 | 776.5 | 753.5 | IC | IC | IC | IC |
| 20.0 | 856.8 | 840.8 | 806.0 | 788.1 | 769.2 | 730.4 | IC | IC |
| 30.0 | 872.9 | 858.7 | 828.6 | 813.2 | 797.1 | 759.4 | 729.7 | IC |
| 40.0 | 886.9 | 874.3 | 847.5 | 833.7 | 819.3 | 783.6 | 758.0 | 729.3 |
| 50.0 | 899.4 | 888.1 | 863.6 | 850.8 | 837.7 | 804.9 | 781.2 | 756.0 |
| 60.0 | 911.1 | 900.6 | 877.8 | 865.9 | 853.6 | 823.5 | 801.3 | 778.2 |
| 70.0 | 922.0 | 911.9 | 890.5 | 879.2 | 867.6 | 839.9 | 818.9 | 797.2 |
| 80.0 | 932.3 | 922.6 | 902.0 | 891.3 | 880.3 | 854.6 | 834.4 | 814.1 |
| 90.0 | 942.0 | 932.3 | 912.6 | 902.3 | 891.9 | 867.5 | 848.6 | 829.0 |
| 100.0 | 950.6 | 941.4 | 922.5 | 912.7 | 902.4 | 879.6 | 861.1 | 842.5 |
| | | | | $x_1 = 0.9517 \pm 0.0010$ | | | | |
| 10.0 | 892.7 | 851.6 | IC | IC | IC | IC | IC | IC |
| 20.0 | 933.2 | 902.4 | 834.7 | 796.8 | 769.6 | IC | IC | IC |
| 30.0 | 963.2 | 937.2 | 882.0 | 852.8 | 819.5 | 765.5 | IC | IC |
| 40.0 | 987.6 | 964.4 | 916.2 | 891.3 | 864.5 | 813.5 | 763.7 | IC |
| 50.0 | 1008.3 | 987.0 | 943.4 | 920.9 | 899.8 | 856.4 | 810.7 | 765.5 |
| 60.0 | 1026.0 | 1006.3 | 966.1 | 945.4 | 928.9 | 891.5 | 848.7 | 806.2 |
| 70.0 | 1041.8 | 1023.2 | 985.5 | 966.4 | 953.1 | 920.3 | 879.5 | 839.9 |
| 80.0 | 1056.1 | 1038.6 | 1002.8 | 984.8 | 974.3 | 944.0 | 905.2 | 867.7 |
| 90.0 | 1069.2 | 1052.5 | 1018.6 | 1001.3 | 993.2 | 963.5 | 927.8 | 892.0 |
| 100.0 | 1081.2 | 1065.2 | 1032.7 | 1016.2 | 1009.8 | 979.0 | 943.6 | 912.6 |

[a] Standard uncertainties ($k$ = 1): $u_r(p) = 2 \times 10^{-4}$ (Pa/Pa); $u(T)$ = 10 mK, $u(p)$ = 0.9 kg m$^{-3}$.

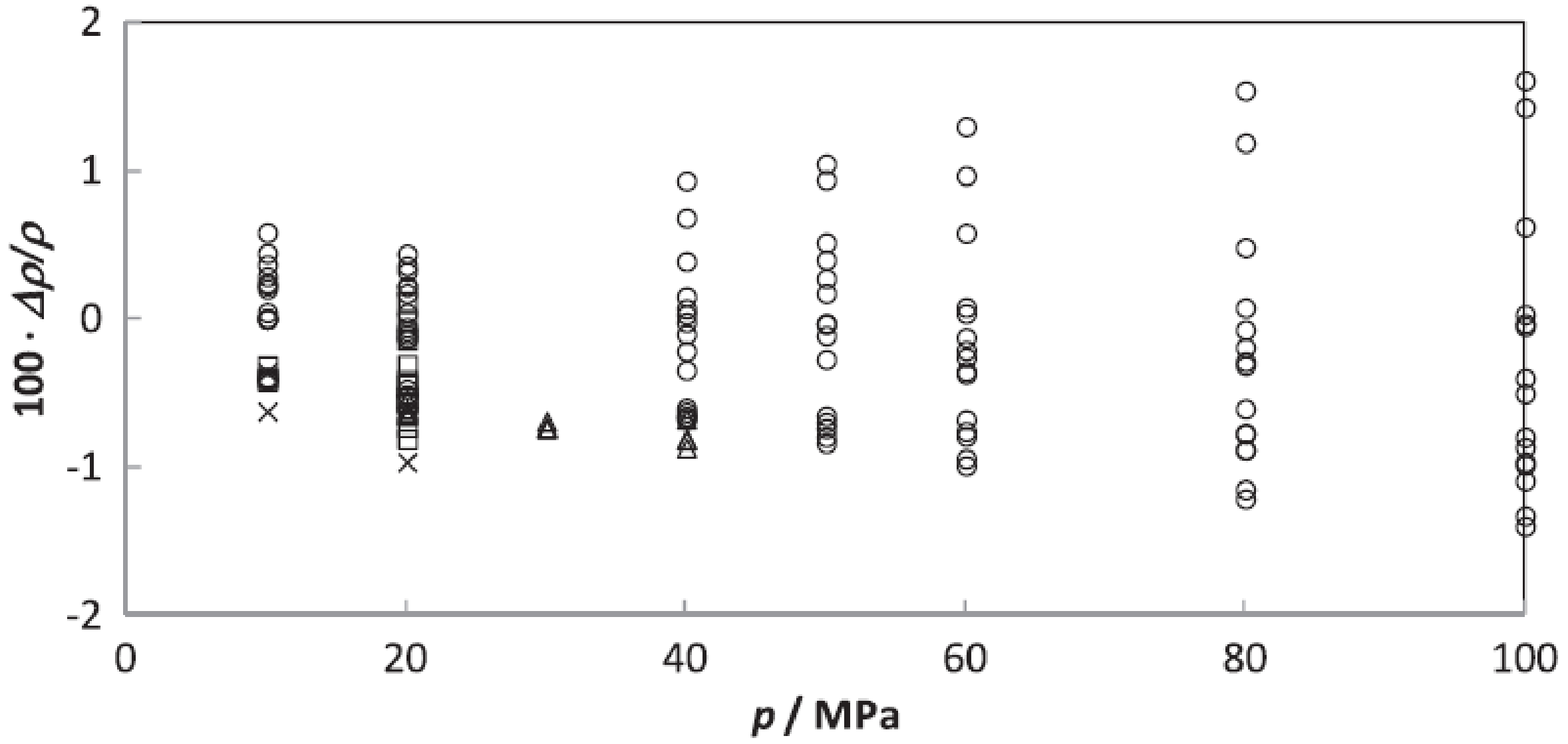

**Fig. 2.** Relative deviation of density values of the (CO2 + *n*-decane) mixture from the literature as a function of pressure: (O) Fandiño et al. [10]; (□), Zuñiga et al. [11]; (×) Cullick et al. [17] and (Δ) Bessieres et al. [18].

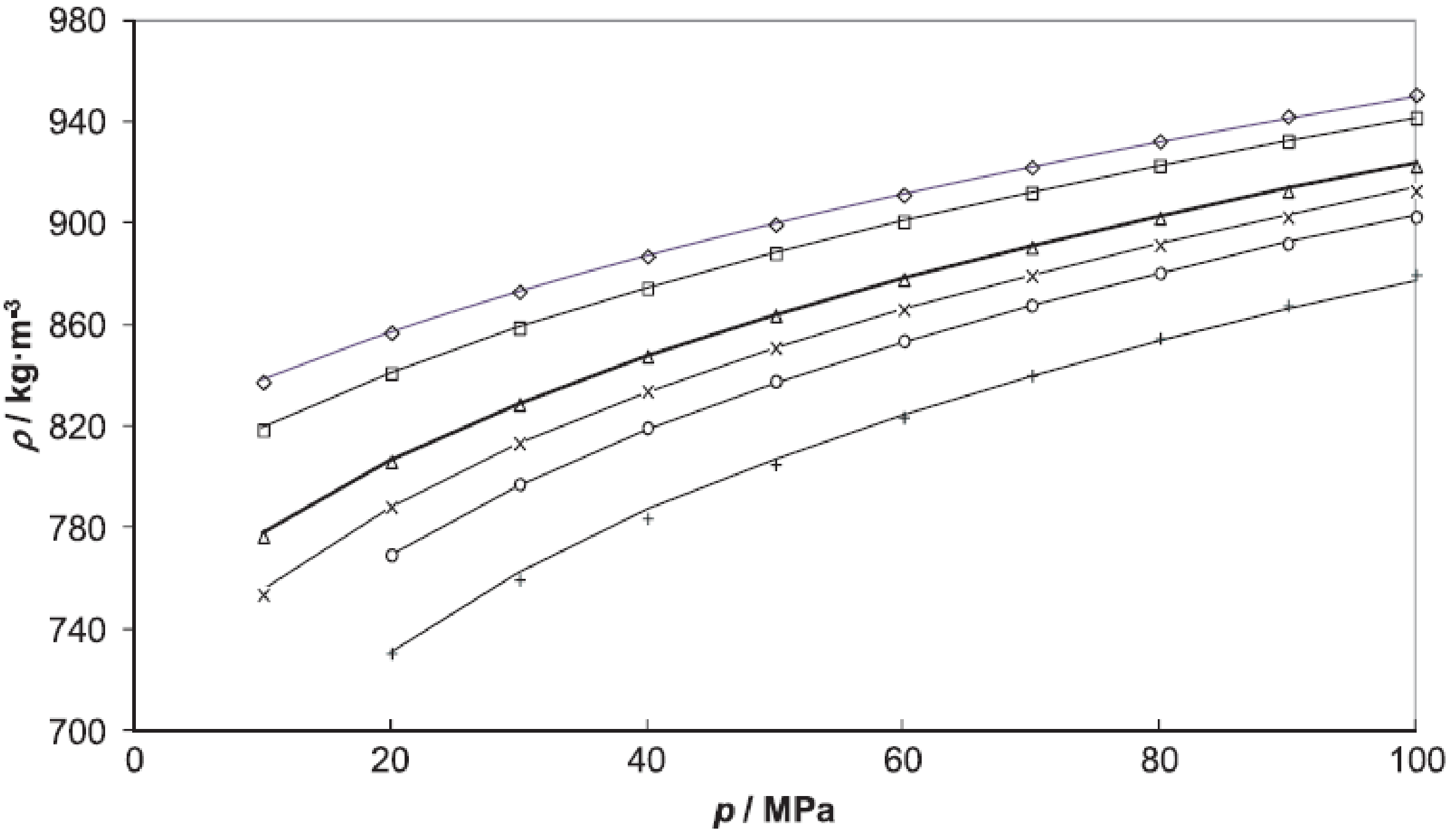

**Fig. 3.** Density of the $CO_2$ ($x_1$) + *n*-decane ($1 - x_1$) mixture, $x_1 = 0.7970$, as a function of pressure at different temperatures: ◊ 283.15 K, □ 293.15 K, Δ 313.15 K, × 323.15 K, O 333.15 K, + 353.15 K. The line represents the calculated values using Tamman–Tait equation.

In Fig. 4, the behaviour for the {($x_1$) $CO_2$ + ($1 - x_1$) *n*-decane} mixture is observed as a function of composition at different pressures and supercritical conditions for $CO_2$ ($T > 304.1$ K and $p > 7.38$ MPa). As expected, densities are higher affected by pressure changes when $CO_2$ composition is increased.

The results obtained for this system validate the experimental technique, and the study continues measuring ($CO_2$ + *n*-dodecane) and ($CO_2$ + squalene) mixtures.

The experimental densities for the system {($x_1$) $CO_2$ + ($1 - x_1$) *n*-dodecane} at four compositions $x_1$ = (0.1014, 0.1998, 0.4001 and 0.6022) are reported in Table 4.

Density behaviour for the mixture {($x_1$) $CO_2$ + ($1 - x_1$) *n*- dodecane}, is plotted against pressure, at various temperatures and $x_1 = 0.6022$, in Fig. 5 and, as a function of composition at $T = 313.15$ K and different pressures, in Fig. 6. Immiscibility was observed at $x_1 = 0.6022$, $T = 393.15$ K and $p = 10$ MPa.

As expected, densities increase when pressure is increased and temperature is decreased; the plot also shows that these effects are greater at the highest temperature and the lowest pressure.

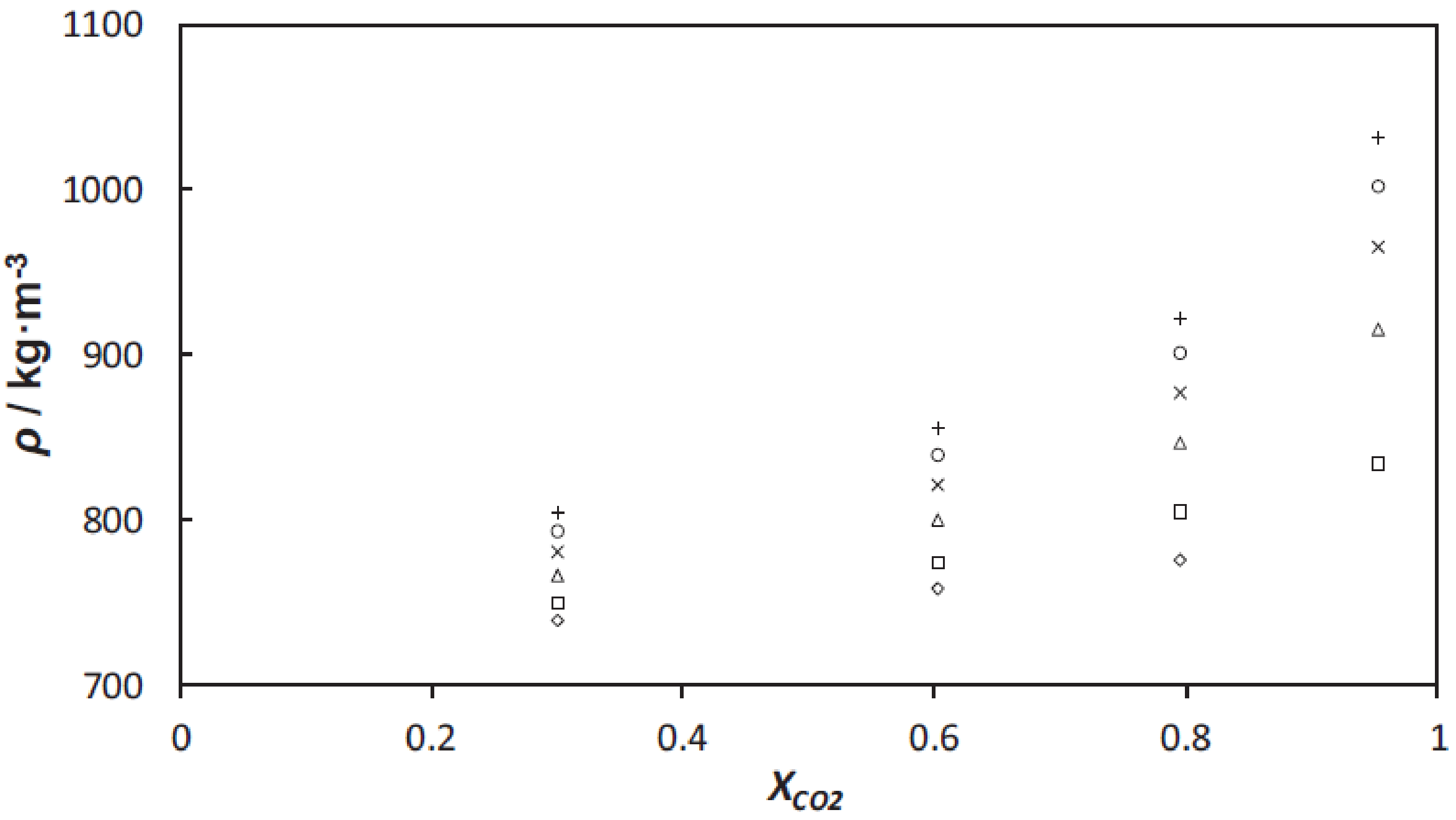

**Fig. 4.** Density of the {($x_1$) $CO_2$ + (1 − $x_1$) *n*-decane} mixture at $T$ = 313.15 K as a function of the composition at different pressures: ◊ 10 MPa, □ 20 MPa, Δ 40 MPa, × 60 MPa, O 80 MPa, + 100 MPa.

**Table 4.** Experimental densities, $\rho$ (kg m$^{-3}$), for the {($x_1$) $CO_2$ + (1 − $x_1$) *n*-dodecane} mixture, at different pressures and temperatures[a]. IC stands for inmiscibility conditions.

| *T*/K | 283.15 | 293.15 | 313.15 | 323.15 | 333.15 | 353.15 | 373.15 | 393.15 |
|---|---|---|---|---|---|---|---|---|
| *p*/MPa | | | | $\rho$/(kg m$^{-3}$) | | | | |
| | | | | $x_1$ = 0.1014 ± 0.0017 | | | | |
| 10.0 | 762.1 | 755.4 | 741.8 | 734.9 | 728.2 | 714.5 | 700.5 | 686.8 |
| 20.0 | 768.4 | 762.0 | 749.0 | 742.5 | 736.3 | 723.4 | 710.4 | 697.7 |
| 30.0 | 774.2 | 768.0 | 755.6 | 749.4 | 743.5 | 731.4 | 719.5 | 707.5 |
| 40.0 | 779.6 | 773.6 | 761.6 | 755.9 | 750.4 | 739.0 | 727.8 | 716.5 |
| 50.0 | 784.7 | 778.9 | 767.4 | 761.8 | 756.8 | 746.1 | 735.5 | 724.6 |
| 60.0 | 789.6 | 784.0 | 772.8 | 767.6 | 763.0 | 752.7 | 742.8 | 732.4 |
| 70.0 | 794.1 | 788.7 | 777.9 | 773.1 | 769.0 | 759.4 | 749.8 | 739.7 |
| 80.0 | 798.5 | 793.3 | 782.9 | 778.6 | 774.9 | 765.8 | 756.7 | 747.0 |
| 90.0 | 803.0 | 797.7 | 787.8 | 784.1 | 780.8 | 771.9 | 763.0 | 753.6 |
| 100.0 | 806.9 | 802.0 | 792.7 | 789.4 | 786.3 | 777.9 | 769.3 | 760.2 |
| | | | | $x_1$ = 0.1998 ± 0.0021 | | | | |
| 10.0 | 762.3 | 755.5 | 741.8 | 735.1 | 728.6 | 715.2 | 701.3 | 687.1 |
| 20.0 | 768.6 | 762.1 | 749.0 | 742.9 | 737.0 | 724.5 | 711.7 | 698.6 |
| 30.0 | 774.4 | 768.1 | 755.7 | 750.0 | 744.7 | 733.0 | 721.2 | 708.8 |
| 40.0 | 779.8 | 773.7 | 761.9 | 756.8 | 751.9 | 741.0 | 730.1 | 718.2 |
| 50.0 | 784.9 | 779.1 | 767.7 | 763.2 | 758.8 | 749.4 | 738.2 | 726.9 |
| 60.0 | 789.8 | 784.2 | 773.3 | 769.6 | 765.6 | 756.5 | 745.9 | 735.1 |
| 70.0 | 794.4 | 788.9 | 778.8 | 775.7 | 772.0 | 763.5 | 753.2 | 742.7 |
| 80.0 | 798.8 | 793.6 | 784.5 | 781.8 | 778.3 | 770.3 | 760.1 | 750.6 |
| 90.0 | 803.2 | 798.3 | 790.0 | 787.6 | 784.3 | 776.6 | 767.1 | 757.6 |
| 100.0 | 807.2 | 802.8 | 795.6 | 793.3 | 790.4 | 782.6 | 773.4 | 764.1 |
| $x_1$ = 0.4001 ± 0.0029 | | | | | | | | |
| 10.0 | 762.7 | 755.8 | 742.1 | 735.6 | 729.2 | 715.7 | 701.4 | 687.1 |
| 20.0 | 768.9 | 762.5 | 749.6 | 743.6 | 737.7 | 725.2 | 711.8 | 698.7 |
| 30.0 | 774.7 | 768.7 | 756.4 | 750.8 | 745.6 | 733.8 | 721.5 | 709.0 |
| 40.0 | 780.2 | 774.6 | 762.8 | 757.7 | 752.9 | 742.0 | 730.5 | 718.6 |
| 50.0 | 785.6 | 780.2 | 768.8 | 764.3 | 760.1 | 749.6 | 738.7 | 727.3 |
| 60.0 | 790.7 | 785.6 | 774.7 | 770.8 | 767.0 | 756.9 | 746.5 | 735.5 |
| 70.0 | 795.7 | 790.7 | 780.5 | 777.1 | 773.6 | 764.0 | 754.0 | 743.3 |
| 80.0 | 800.6 | 795.8 | 786.4 | 783.2 | 780.1 | 770.9 | 761.3 | 750.9 |
| 90.0 | 805.5 | 800.7 | 791.9 | 789.2 | 786.2 | 777.4 | 767.8 | 758.0 |
| 100.0 | 810.2 | 805.6 | 797.5 | 794.9 | 792.2 | 783.4 | 774.4 | 764.8 |
| $x_1$ = 0.6022 ± 0.0028 | | | | | | | | |
| 10.0 | 775.2 | 768.8 | 753.4 | 746.0 | 738.3 | 721.9 | 704.4 | IC |
| 20.0 | 782.8 | 777.7 | 763.2 | 756.4 | 749.4 | 734.0 | 717.8 | 701.9 |
| 30.0 | 790.1 | 785.8 | 772.0 | 765.9 | 759.4 | 744.8 | 729.9 | 714.9 |
| 40.0 | 797.5 | 793.5 | 780.3 | 774.8 | 768.6 | 754.9 | 740.9 | 726.6 |
| 50.0 | 804.3 | 800.5 | 788.2 | 783.0 | 777.3 | 764.1 | 750.7 | 737.1 |
| 60.0 | 811.5 | 807.3 | 795.7 | 790.8 | 785.4 | 772.7 | 759.9 | 746.9 |
| 70.0 | 817.8 | 813.6 | 802.9 | 798.2 | 792.9 | 780.9 | 768.4 | 756.0 |
| 80.0 | 824.2 | 819.9 | 809.8 | 805.3 | 800.2 | 788.8 | 776.5 | 764.6 |
| 90.0 | 830.6 | 826.1 | 816.4 | 812.0 | 807.1 | 795.9 | 784.3 | 772.7 |
| 100.0 | 836.9 | 831.8 | 822.7 | 818.4 | 813.7 | 802.7 | 791.5 | 780.2 |

[a] Standard uncertainties ($k$ = 1): $u_r(p) = 2 \times 10^{-4}$ (Pa/Pa); $u(T)$ = 10 mK, $u(\rho)$ = 0.9 kg m$^{-3}$.

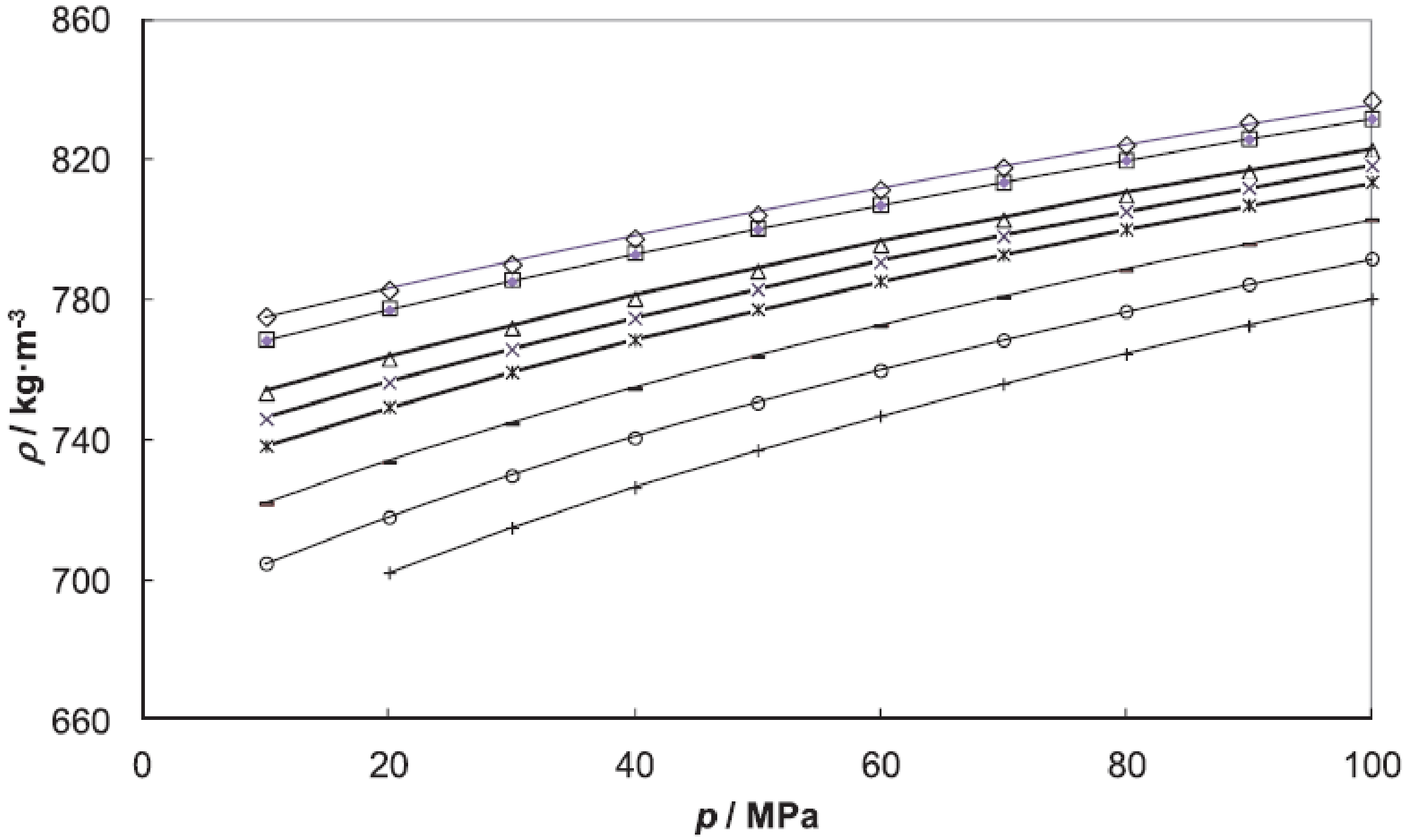

**Fig. 5.** Densities of the {($x_1$) $CO_2$ + (1 − $x_1$) *n*-dodecane} mixture, $x_1$ = 0.6022, as a function of pressure at different temperatures: ◊ 283.15 K, □ 293.15 K, Δ 313.15 K, × 323.15 K, ✳ 333.15 K, – 353.15 K, O 373.15 K, + 393.15 K. The line represents the calculated values using Tamman–Tait equation.

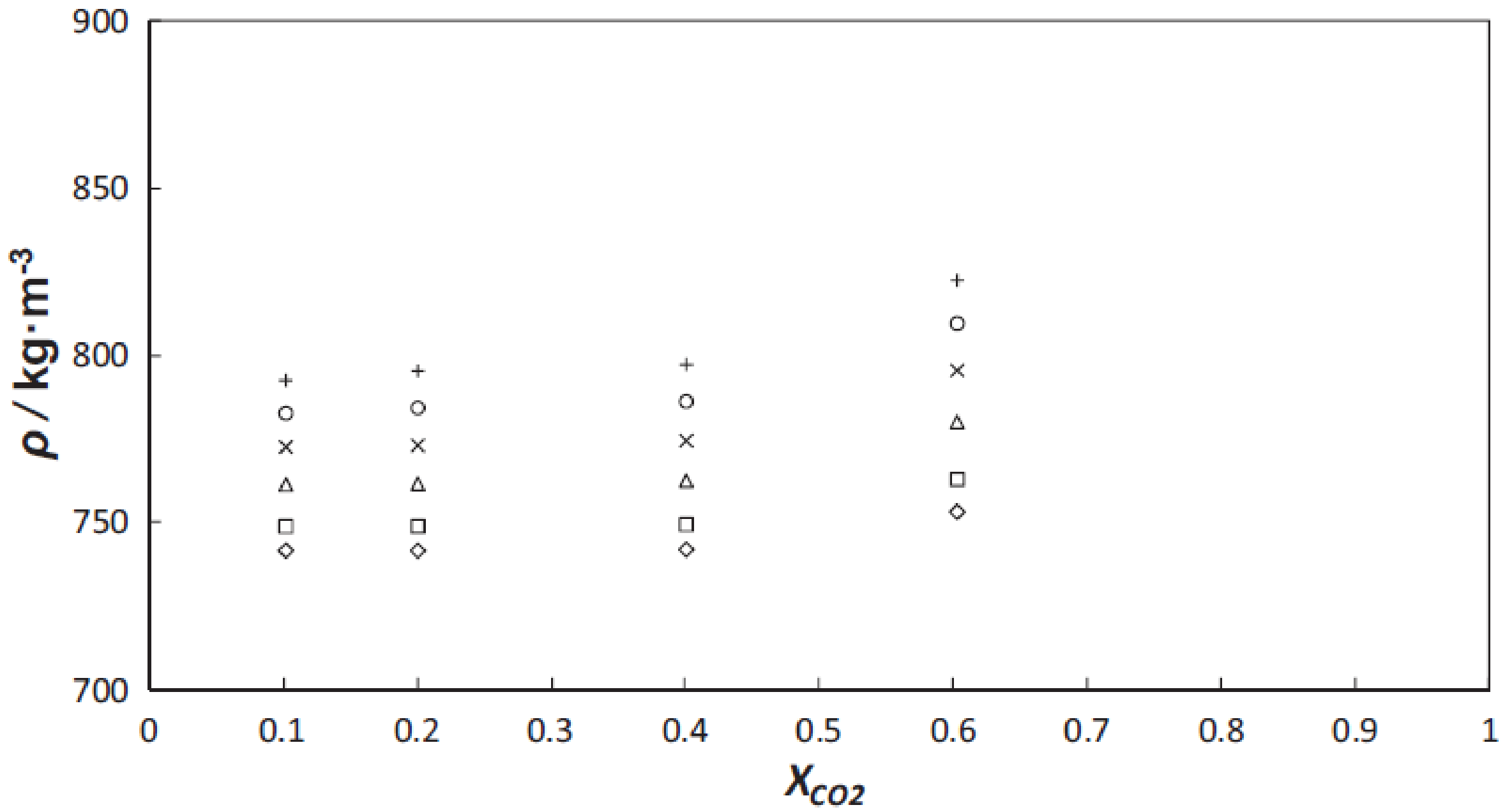

**Fig. 6.** Densities of the {($x_1$) $CO_2$ + (1 − $x_1$) *n*-dodecane} mixture as a function of the composition at $T$ = 313.15 K and different pressures: ◊ 10 MPa, □ 20 MPa, Δ 40 MPa, × 60 MPa, O 80 MPa, + 100 MPa.

Moreover, Fig. 6 shows the reduced dependence of the density on $CO_2$ mole fraction when this is lower than 0.5. This effect was also described in the literature [19] and may be attributed to the situation of $CO_2$ molecules inside the gaps between dodecane molecules. We only found these data [19] for comparison, however, temperature and pressure ranges are smaller than ours: they reported data for the system ($CO_2$ + dodecane) at $T$ = (313.55–353.55) K and $p$ = (8–18) MPa for four compositions different from ours. The comparison between both sets of data gives differences less than 1.5%, therefore there is a good agreement between them, as seen in Fig. 7 were the excess volumes are plotted as a function of the composition.

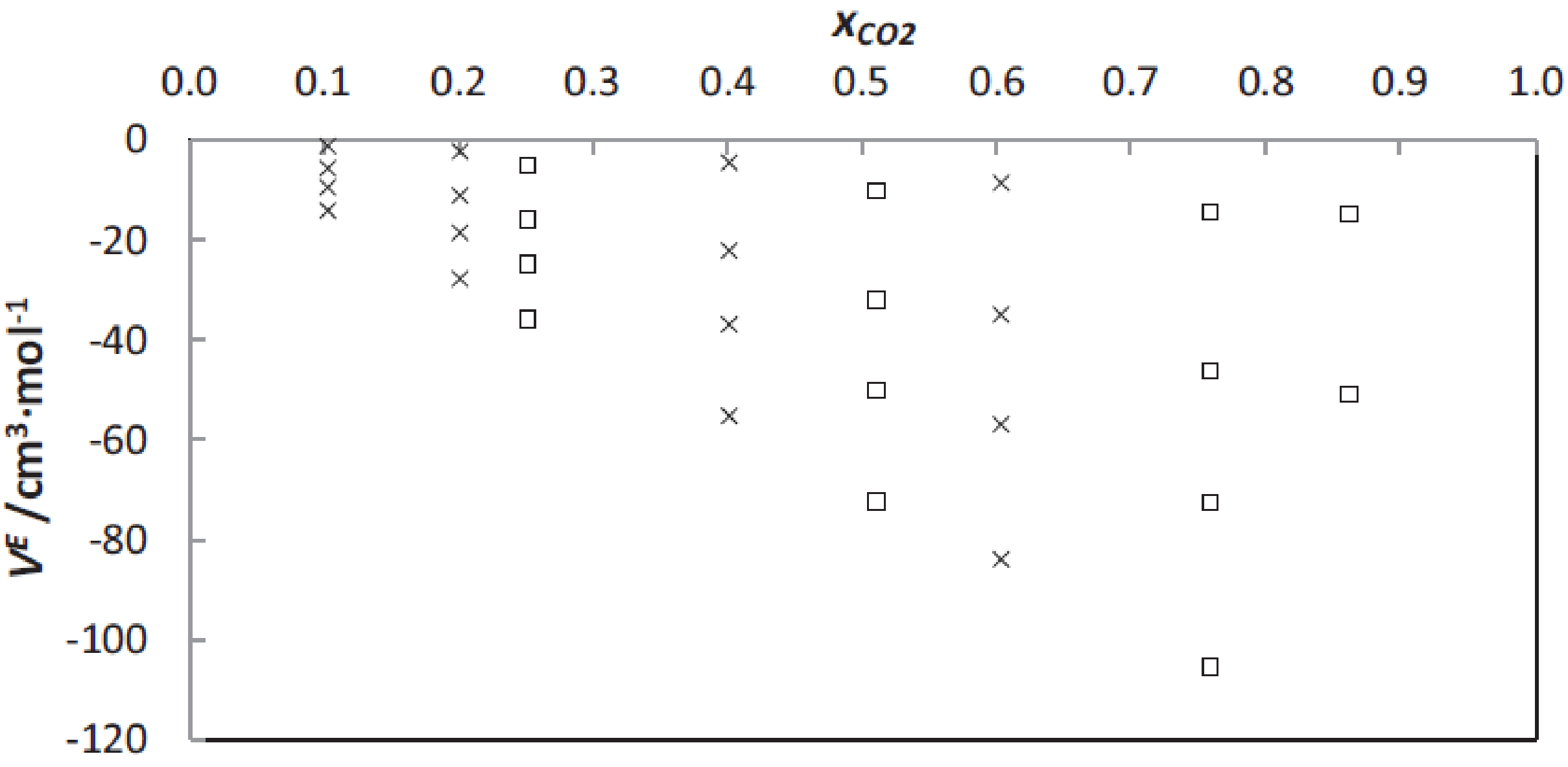

**Fig. 7.** Excess volumes of the ($CO_2$ + *n*-dodecane) mixture at $T$ = (313.15, 323.15, 333.15, 353.15) K and $p$ = 10 MPa in comparison with literature: (×) this work and (□) Zhang et al. [19].

Squalane is a hydrocarbon of 30 carbon atoms and was selected as heavy hydrocarbon [20] in order to model the behaviour of heavy oils. Two mixtures $\{(x_1)\ CO_2 + (1 - x_1)\ \text{squalene}\}$ at $x_1$ = (0.1001 and 0.2012) were measured and their densities are reported in Table 5.

In Fig. 8, the density behaviour is plotted against temperature at mole fraction $x_1$ = 0.2012 for various pressures. Immiscibility for this mixture was obtained at $p$ = 10 MPa, $T$ = (373.15 and 393.15) K and $x_1$ = 0.2012.

**Table 5.** Experimental densities, $\rho$ (kg m$^{-3}$), for the mixtures {($x_1$) $CO_2$ + (1 − $x_1$) squalene} at different pressures and temperatures.[a]

| $T$/K | 283.15 | 293.15 | 313.15 | 323.15 | 333.15 | 353.15 | 373.15 | 393.15 |
|---|---|---|---|---|---|---|---|---|
| $p$/MPa | $\rho$/(kg m$^{-3}$) | | | | | | | |
| | $x_1$ = 0.1001 ± 0.0017 | | | | | | | |
| 10.0 | 823.3 | 816.8 | 804.1 | 797.9 | 791.8 | 779.3 | 766.7 | 754.3 |
| 20.0 | 828.8 | 822.8 | 810.6 | 804.6 | 798.7 | 786.9 | 774.9 | 763.3 |
| 30.0 | 834.1 | 828.2 | 816.4 | 810.7 | 804.9 | 793.7 | 782.5 | 771.3 |
| 40.0 | 839.0 | 833.3 | 822.0 | 816.4 | 811.0 | 800.1 | 789.4 | 778.6 |
| 50.0 | 843.7 | 838.0 | 827.1 | 821.7 | 816.4 | 806.0 | 795.8 | 785.3 |
| 60.0 | 848.2 | 842.7 | 832.0 | 826.8 | 821.6 | 811.4 | 801.5 | 791.5 |
| 70.0 | 852.4 | 847.0 | 836.6 | 831.5 | 826.4 | 816.6 | 807.0 | 797.3 |
| 80.0 | 856.5 | 851.3 | 841.0 | 836.0 | 831.1 | 821.6 | 812.4 | 802.8 |
| 90.0 | 860.4 | 855.3 | 845.3 | 840.5 | 835.7 | 826.3 | 817.2 | 808.1 |
| 100.0 | 864.1 | 859.1 | 849.4 | 844.8 | 839.9 | 830.8 | 821.9 | 813.1 |
| | $x_1$ = 0.2012 ± 0.0021 | | | | | | | |
| 10.0 | 844.5 | 836.0 | 818.8 | 810.2 | 801.6 | 784.2 | | |
| 20.0 | 852.4 | 844.4 | 828.2 | 820.1 | 812.0 | 796.0 | 778.9 | 763.5 |
| 30.0 | 859.6 | 851.9 | 836.6 | 828.9 | 821.3 | 806.2 | 790.3 | 775.5 |
| 40.0 | 866.2 | 858.8 | 844.1 | 836.9 | 829.5 | 815.3 | 801.1 | 786.4 |
| 50.0 | 872.3 | 865.2 | 851.1 | 844.0 | 837.2 | 823.4 | 810.9 | 796.5 |
| 60.0 | 878.1 | 871.1 | 857.5 | 850.8 | 844.2 | 830.9 | 819.9 | 805.7 |
| 70.0 | 883.5 | 876.7 | 863.5 | 857.0 | 850.6 | 838.0 | 828.1 | 814.4 |
| 80.0 | 888.6 | 882.0 | 869.2 | 863.0 | 856.7 | 844.6 | 835.5 | 822.6 |
| 90.0 | 893.5 | 887.1 | 874.6 | 868.5 | 862.5 | 850.8 | 842.4 | 829.9 |
| 100.0 | 898.1 | 891.8 | 879.8 | 873.7 | 867.8 | 856.4 | 849.0 | 836.8 |

[a] Standard uncertainties ($k$ = 1): $u_r(p) = 2 \times 10^{-4}$ (Pa/Pa); $u(T)$ = 10 mK, $u(p)$ = 0.9 kg m$^{-3}$.

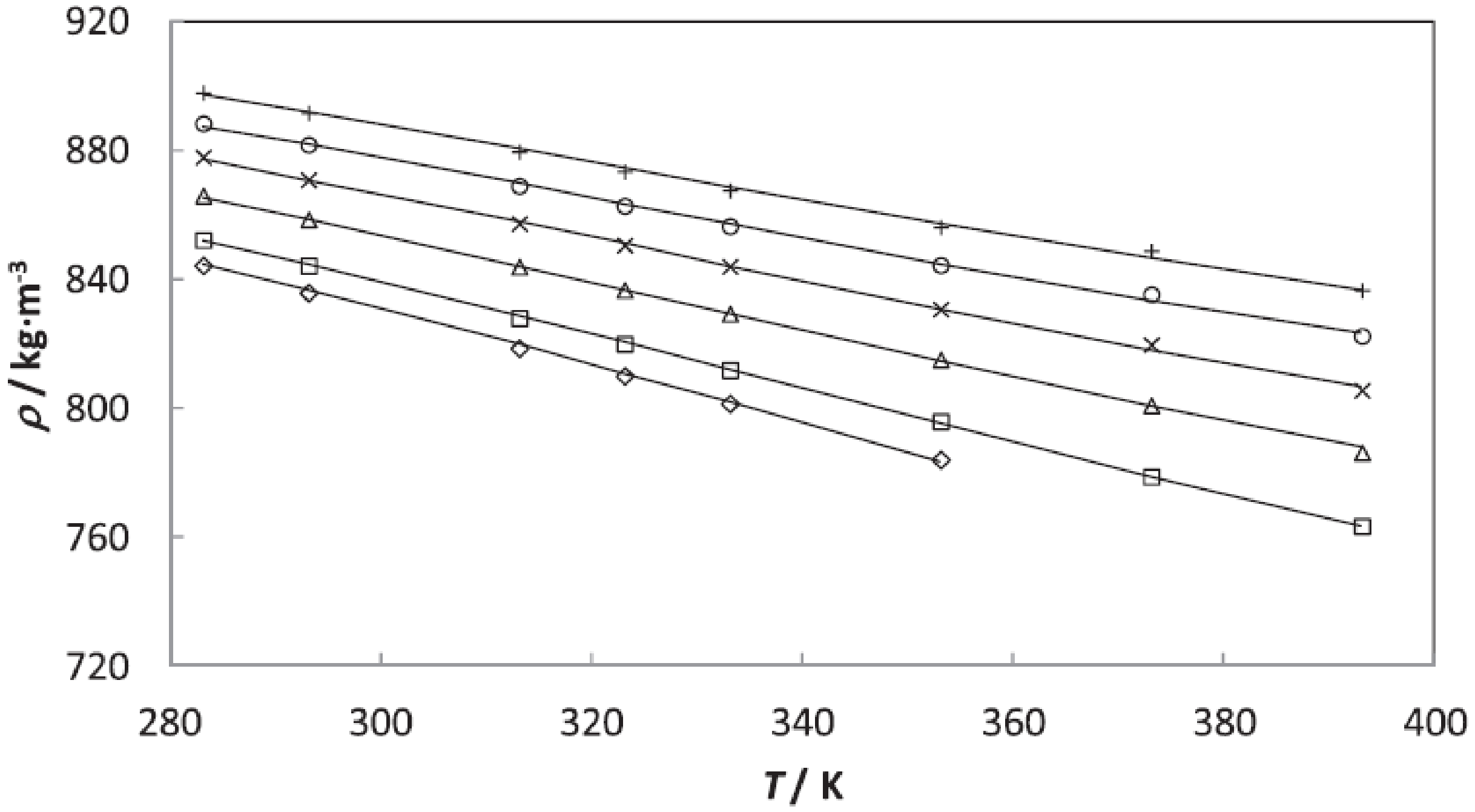

**Fig. 8.** Density of the {($x_1$) $CO_2$ + (1 − $x_1$) squalene} mixture, $x_1$ = 0.2012, as a function of temperature at different pressures: ◊ 10 MPa, □ 20 MPa, Δ 40 MPa, × 60 MPa, O 80 MPa, + 100 MPa. The line represents the calculated values using Tamman–Tait equation.

We only found one reference [21] which also reports experimental densities for this system at three compositions higher than ours at $T$ = (303.15–448.15) K and up to 170 MPa, but their temperatures are also different. Both sets of data are coherent in the sense that both studies show how the pressure dependence of density is more pronounced at lower temperatures.

Finally, the modified Tamman–Tait equation [22] was used to fit the experimental data for all the studied mixtures:

$$\rho(T,p) = \frac{\rho(T,p_{ref})}{1 - C\ln[(B(T)+p)/(B(T)+p_{ref})]} \quad (2)$$

where $C$ is a parameter independent of temperature, $p(T, p_{\mathrm{ref}})$ is the temperature dependence of the density at a reference pressure and B($T$) is a parameter which depends on temperature. The temperature dependence of these parameters is given by the following equations:

$$\rho(T,p_{ref}) = \sum_{i=0}^{N} A_i T^i \quad (3)$$

$$B(T) = \sum_{i=0}^{N} B_i T^i \quad (4)$$

The parameters ($A_i$, $B_i$, $C$) were determined by least squares method, for each composition of the mixtures. The results of the fitting are summarized in Table 6 which also contains the standard deviations.

The values of standard deviations range from 0.33 to 0.86, except for the mixture {$CO_2$ (0.9517) + decane (0.0483)} which is close to pure $CO_2$. Therefore, Eq. (2) reproduces quite well the behaviour of this kind of mixtures.

**Table 6.** Fitting parameters and standard deviations using Eq. (2) for the studied mixtures.

| $x_1$ | $A_0$ (kg m$^{-3}$) | $10^3 \times A_1$ (kg m$^{-3}$ K$^{-1}$) | $10^6 \times A_2$ (kg m$^{-3}$ K$^{-2}$) | $10^9 \times A_3$ (kg m$^{-3}$ K$^{-3}$) | $B_0$ (MPa) | $B_1$ (MPa K$^{-1}$) | $10^3 \times B_2$ (MPa K$^{-2}$) | $C$ | (kg m$^{-3}$) |
|---|---|---|---|---|---|---|---|---|---|
| | | | | $CO_2$ ($x_1$) + decane (1 − $x_1$) | | | | | |
| 0.3011 | 0.6401 | 2.335 | −9.009 | 8.225 | 350.0 | −1.373 | 1.367 | 0.0918 | 0.33[a] |
| 0.6052 | 0.6944 | 1.938 | −6.417 | 2.874 | 405.6 | −1.805 | 1.997 | 0.1028 | 0.58[a] |
| 0.7970 | 0.0352 | 9.966 | −34.27 | 32.93 | 541.4 | −2.870 | 3.791 | 0.1060 | 0.86[b] |
| 0.9517 | 0.4739 | 11.30 | −50.58 | 57.94 | 723.4 | −4.242 | 6.139 | 0.1131 | 2.8[b] |
| $x_1$ | $A_0$ (kg m$^{-3}$) | $10^4 \times A_1$ (kg m$^{-3}$ K$^{-1}$) | $10^7 \times A_2$ (kg m$^{-3}$ K$^{-2}$) | $10^{10} \times A_3$ (kg m$^{-3}$ K$^{-3}$) | $B_0$ (MPa) | $B_1$ (MPa K$^{-1}$) | $10^3 \times B_2$ (MPa K$^{-2}$) | $C$ | (kg m$^{-3}$) |
| | | | | $CO_2$ ($x_1$) + dodecane (1 − $x_1$) | | | | | |
| 0.1014 | 0.9409 | −5.784 | −2.247 | 1.306 | 675.5 | −2.711 | 2.933 | 0.1200 | 0.42[a] |
| 0.1998 | 0.9390 | −5.875 | −1.236 | −0.2743 | 952.7 | −4.232 | 5.034 | 0.1301 | 0.56[a] |
| 0.4001 | 0.9360 | −5.882 | −0.1983 | −2.373 | 852.9 | −3.681 | 4.289 | 0.1332 | 0.51[a] |
| 0.6022 | 0.5593 | 28.64 | −97.99 | 87.46 | 456.2 | −1.775 | 1.853 | 0.1246 | 0.42[b] |
| $x_1$ | $A_0$ (kg m$^{-3}$) | $10^3 \times A_1$ (kg m$^{-3}$ K$^{-1}$) | $10^5 \times A_2$ (kg m$^{-3}$ K$^{-2}$) | $10^8 \times A_3$ (kg m$^{-3}$ K$^{-3}$) | $B_0$ (MPa) | $B_1$ (MPa K$^{-1}$) | $10^3 \times B_2$ (MPa K$^{-2}$) | $10^2 \times C$ | (kg m$^{-3}$) |
| | | | | $CO_2$ ($x_1$) + squalane (1 − $x_1$) | | | | | |
| 0.1001 | 0.6742 | 2.332 | −0.8880 | 0.8823 | 394.9 | −1.356 | 1.272 | 8.598 | 0.37[a] |
| 0.2012 | 0.5935 | 3.586 | −1.310 | 1.291 | 447.0 | −1.759 | 1.788 | 9.399 | 0.84[b] |

a $p_{ref}$ = 10 MPa.

b $p_{ref}$ = 20 MPa.

## 4. Conclusions

Experimental equipment was implemented for preparation and density determination of ($CO_2$ + hydrocarbon) mixtures. Mixtures are prepared using flow pumps in such a way that ensures phase homogenization.

The technique was validated using the mixture ($CO_2$ + *n*-decane) as test system obtaining an average relative deviation from the literature of 0.5%.

Four mixtures of ($CO_2$ + *n*-dodecane) and two mixtures of ($CO_2$ + squalene) were measured ranging from 283.15 K to 393.15 K (eight isotherms) and 10–100 MPa (ten isobars). The immiscibility of the blends is detected by the instability of the period readings in the unit of measurement of the vibrating-tube densimeter. This condition for the mixtures ($CO_2$ + hydrocarbon) appears at high temperatures and low pressures as was detailed above.

**Acknowledgments**

J. Zambrano thanks to TERMOCAL Research Group of the Uni versity of Valladolid, and to the Secretaría Nacional de Educación Superior, Ciencia, Tecnología e Innovación (SENESCYT), Ecuador, through a scholarship for doctoral studies. The work was funded by the Regional Government of Castilla y León through the Project VA295U14.